\documentclass{elsart}

\usepackage{amsmath}
\begin{document}
\begin{frontmatter}
\title{Reconsidering the black hole final state in Dirac fields}
\author[shao,nao,gs]{Xian-Hui Ge}
\author[shao,nao,itp,email]{You-Gen Shen}
\address[shao]{Shanghai Astronomical Observatory,Chinese Academy of Sciences,
Shanghai 200030, China (post address)}
\address[nao]{National Astronomical Observatories, Chinese Academy of Sciences,
Beijing 100012, China}
\address[gs]{Graduate School of Chinese
Academy of Sciences, Beijing 100039, China}
\address[itp]{Institute of Theoretical Physics, Chinese Academy of Sciences,
Beijing 100080, China}
\thanks[email]{e-mail:gexh@shao.ac.cn\\ ygshen@shao.ac.cn}

\begin{abstract}
  We extend Horowitz and Maldacena's proposal about black hole final state to Dirac fields and find
  that if annihilation of the infalling positrons and the collapsed electrons inside the horizon is
  considered, then the nonlinear evolution of collapsing quantum state will be
  avoided. We further propose that annihilation also plays the
  central role in the process of black hole information escaping
  in both Dirac and scalar fields.
  The computation speed of a black hole is also briefly discussed.
\end{abstract}
\end{frontmatter}
\newpage
\hspace*{7.5mm}The information loss of a black hole has long been
a puzzling problem
 to most theoretical physicists[1]. It has been considered as one of the most
challenging problems in the 21st century theoretical physics.
Several schools of thoughts have emerged for solving the black
hole information paradox[2,3,4,5]. Nowadays, evidence in string
theory comes to support that the formation and evaporation of
black holes is a unitary process[6,7,8,9]. But semiclassical
arguments suggest that the black hole evaporation process is not
unitary. Therefore, if there is no information loss, it remains a
mystery how information is returned in the semiclassical process.
In other words, we are confronted with a challenge of reconciling
the semiclassical reasoning with the putative unitary of black
hole evaporation. On the other hand, while people are puzzled by
the black hole information paradox, quantum mechanics, especially
the new field of quantum information, has undergone rapid progress
theoretically and experimentally, as the consequence of the
further development of the debate between Einstein and Bohr on the
completeness of quantum theory. The theoretical and experimental
realization of quantum teleportation suggests that entanglement
plays a central role in the quantum teleportation and quantum
computation[10]. Although, up to now, no experiment can clearly
tell who (Einstein or Bohr) is completely correct, the concept of
entanglement, which stems from the 'Einstein-Podolsky-Rosen (EPR)
paradox' [11], has shown its unique properties without any
classical correspondence. Evidence has shown that entanglement is
closely related to the effects of gravity: quantum fluctuations
near a black hole horizon is filled with pairs of entangled
particles and anti-particles[12,13,14]; One of the particles flies
outward to become the Hawking radiation that an observer sees; The
other falls into the horizon. This phenomena implies scientists
that a black hole might be a huge
computer system[15,16,17].\\
\hspace*{7.5mm} Horowitz and Maldacena (H-M) have come to be the
first to realize black hole computer in theory[18]. They have
proposed a simple model of black hole evaporation by imposing a
final boundary condition between string theory and semiclassical
argument over whether black hole evaporation is unitary. This
model requires a specific final state at black hole singularity
which is entangled between the collapsing matter and the incoming
Hawking radiation. However, Gottsman, Preskill, and later
Yurtsever and Hockney argued that the proposed constraint must
lead to nonlinear evolution of the initial quantum state, and one
cannot ensure the black hole
final state to be maximally entangled[19,20].\\
\hspace*{7.5mm}Considering the recent teleportation protocol
suggested by Beenaker and Kindermann, which asserts that the
annihilation of a particle-hole pair in the Fermi sea teleports
quasiparticles to distant location, if entanglement was
established beforehand[21], we extend the H-M scenario to Dirac
fields and find that if energy conservation is considered, one can
be free from the worry about nonlinear evolution of the initial
quantum state. From the energy conservation point of view, the
mass of the residual hole must go down. Actually, Hawking
radiation can be described as tunneling. The energy of a particle
changes sign as it crosses the horizon, thus a pair created just
inside or just outside the horizon can be materialize with zero
total energy, after one number of the pair has tunneled to the
opposite side[22]. The annihilation of the infalling particles and
the matter (or the initial quantum state ) can acts as a
measurement, transferring the information contained in the matter
to the outgoing Hawking radiation. In the following, we will take
a Dirac field in the Schwarzschild space-time as an example. It is
well appreciated that the quantization of Dirac fields in
Minkowski and Schwarzscild coordinates are inequivalent[23]. The
Minskowski vacuum will evolve into a state, called the Unruh
state, which can be formulated as
\begin{equation}
\mid0>=N_{f}e^{(itanr
a_{p}^{(\sigma)\dag}a_{\tilde{p}}^{(-\sigma)\dag})}\mid0>_{in}\otimes\mid0>_{out},\end{equation}
where $N_{f}=\cos r$ is the fermionic normalization factor,
$a_{p}^{(\sigma)\dag}$,$a_{p}^{(\sigma)}$,
$a_{\tilde{p}}^{(-\sigma)\dag}$ and $a_{\tilde{p}}^{(-\sigma)}$
are the fermionic creation and annihilation operators inside or
outside the horizon, $p=(\Omega,\vec{k})$ ($\tilde{p}=-p$) is the
mode, r is defined through $tanr=e^{-2\pi M\Omega}$, and the
symbol $\sigma=\pm$ refers to regions inside and outside the
horizon respectively. In the semiclassical approximation, the
overall Hilbert space for the evaporation process of a black hole
can be separated into: $H=H_{M}\otimes H_{in}\otimes H_{out}$,
where $H_{M}$ denotes the Hilbert space of the quantum field that
constitutes the collapsing body, and $H_{in}$ and $H_{out}$
correspond to two parts of the Hilbert space of the fluctuations
which contain modes confined inside and outside the event horizon,
respectively. Considering the finite number of allowed excitations
allowed in fermionic system due to the Pauli exclusion principle,
absorbing factors of i into the definition of the Fock states,
Eq.(1) can rewritten as
\begin{eqnarray}
\mid0>&&=N_{f}\{\mid0>_{in}\otimes\mid0>_{out}+
\sum_{p}tanr\left(\mid 1_{p}>_{in}\right.\nonumber\\
&&\left.\otimes\mid 1_{\tilde{p}}>_{out}+\mid
1_{\tilde{p}}>_{in}\otimes\mid 1_{p}>_{out}\right)\}\end{eqnarray}
Since we only need to consider the information teleported out of
the black hole, we can drop the second set of parentheses in
Eq.(2). We now simplify our analysis by considering the effect of
teleportaiton of the initial quantum state inside the horizon to a
single mode of the outside Hawking particles, which goes as
\begin{equation}
\mid0>=\cos r \mid0>_{in}\otimes\mid0>_{out}+
sinr\mid1>_{in}\otimes\mid1>_{out}. \end{equation}
\hspace*{7.5mm}Hereafter, let us briefly review the H-M proposal
and see where the nonlinear evolution might be raised (For a
detailed and clear review about H-M proposal, please see
Ref.[20]). According to the proposal of H-M, the initial wave
function of collapsing matter $|\Phi>_{M}$ should not be destroyed
inside the horizon and boundary condition on the final quantum
state at the black hole singularity, given by a state called
$<BH|$, which lives in the Hilbert space describing the inside of
the black hole, $H_{M}\otimes H_{in}$, is equal to [18]
\begin{equation}<BH|=\frac{1}{\sqrt{N}}\sum_{m,i}S_{m,i}<j_{M}|\otimes<j_{in}|,\end{equation}
where $\frac{1}{\sqrt{N}}$ is the normalization factor and
$S_{m,i}$ corresponds to the absence of entangling interactions
between $H_{M}$ and ${H_{in}}$:
\begin{equation}
S_{m,i}=S_{m}\otimes S_{i},\end{equation}  which is a unitary
tansformation. Before black hole evaporation, the initial matter
state $|\Phi>_{M}$ evolves into a state in $H_{M}\otimes
H_{in}\otimes H_{out}$, which is given by
$|\Phi>_{M}\otimes|U>_{in\otimes out}$, where $|U>_{in\otimes
out}$ is the Unruh vacuum state in scalar fields given by
\begin{equation}
|U>_{in\otimes out}=\frac{1}{\sqrt{N}}\sum_{k=1}^{N}\mid
k_{in}>\otimes\mid k_{out}>,\end{equation} where $|k_{in}>$ and
$|k_{out}>$ are fixed orthonormal bases for $H_{in}$ and $H_{out}$
respectively. Thus, the outgoing state evolved after the
evaporation can be formulated as\begin{eqnarray}
|Z_{out}>=\frac{\gamma}{N}\sum_{m,i}<j_{M}|\otimes<j_{in}|(S_{m}\otimes
S_{i})\nonumber\\|\Phi>_{M}\otimes
\sum_{k=1}^{N}|k_{in}>\otimes|k_{out}>,
\end{eqnarray}
where $\gamma$ is the renormalization constant that one can define
it to be $\gamma=N$. Assume $Q_{out j}=<j_{out}|X_{out}>$,
$\Phi_{k}=<k_{M}|\Phi_{M}>$,$S_{m jk}=<j_{M}\mid S_{m}\mid
k_{M}>$, and$S_{i jk}=<j_{in}\mid S_{i}\mid k_{in}>$, Eq.(7) can
be rewritten as
\begin{equation}
Z_{out
k}=\frac{\gamma}{N}\sum_{n}(S_{i}^{T}S_{m})_{kn}\Phi_{n},\end{equation}
where $S_{i}^{T}$ denotes matrix transpose of $S_{i}$. Eq.(8)
shows that the tansformation from $H_{M}$ to $H_{out}$ is unitary.
However, as pointed out by Gottsman and Preskill[19], and later by
Yurtsever and Hockney[20], one cannot expect the unitary operator
$S_{mi}$ to have the product form of Eq.(5). The general linear
expression of $Z_{out k}$ should be
\begin{equation}Z_{out k}=\gamma\sum_{n}T_{k n}\Phi_{n},\end{equation}
where T denotes the matrix
\begin{equation}T_{kn}=\frac{1}{\sqrt{N}}<BH\mid n_{M}>\otimes \mid k_{in}>\end{equation}
Only $S_{mi}$ has the product form Eq.(5) T equals a unitary
matrix. Actually, after renormalization, Eq.(9) has a more
succinctly form[20]
\begin{equation}Z_{out k}=\frac{1}{(\sum_{i}|\sum_{j}T_{ij}\Phi_{j}|\mid^2)^{1/2}}\sum_{n}T_{k n}\Phi_{n},\end{equation}
which is not only nonunitary, but it is in fact nonlinear,
linearity along with unitary is recovered only when T is
proportional to a unitary matrix[19,20].\\
\hspace*{7.5mm} However, we would like to point out that the
annihilation of the infalling particles and the collapsed matter
will ensure the maximal entanglement. For example, in Dirac
fields, the positron-electron annihilation is actually
interaction-free and does not reduce the degree of entanglement of
the black hole final state Eq.(4) ![21]. We therefore expect that
there is a final state boundary condition at the singularity,
which demands that the infalling matter and initial quantum state
to be fully annihilated! Without loss of generality, we chose the
initial collapsing matter $|\varphi>_{M}$ in Dirac fields to be
\begin{equation}|\varphi>_{Mi}= \alpha|0>_{Mi}+\beta|1>_{Mi},
\end{equation}which is an unknown state. In fact, the evaporation
of a massive black hole is a long time process, which means that
the total state of the collapsing matter $|\Phi>_{M}$ can be
decomposed into $|\varphi>_{M1},...|\varphi>_{Mi},...$ and every
component will be teleported out in the end. The resulting
annihilation of the infalling positron and the collapsing matter,
say a electron, collapses the combined state
\begin{eqnarray}
&&\left(\alpha|0>_{Mi}+\beta|1>_{Mi}\right)\left(\cos r
\mid0>_{in}\otimes\mid0>_{out}\right
 . \nonumber\\
 &&\left . +
sinr\mid1>_{in}\otimes\mid1>_{out}\right),\end{eqnarray} to the
state $\alpha \cos r\mid0>_{out}+\beta\sin r\mid1>_{out}$ (where
$|0>_{Mi}$ is annihilated with $\mid0>_{in}$ and $|1>_{Mi}$ is
annihilated with $\mid1>_{in}$), which can be further defined as
$\alpha\mid0>_{out}+\beta\mid1>_{out}$, so the state of the matter
is teleported to the outgoing Hawking radiation and the unitarity
is preserved. The similar conclusion can be obtained for the
scalar fields, the annihilation of the infalling matter and the
initially collapsed matter is able to ensure Eq.(11) not to occur.
In this scenario, the uniqueness of the outcomes is still needed,
otherwise instantaneous transfer of information is possible[21].
Since the positron and the collapsing matter (the electron) have
to be annihilated in order to be teleported, no information can be
copied. In contrast to another solution, the black hole
complementarity[5], proposed for the black hole information
problem, the annihilation of the infalling particles and the
collapsing matter present a more natural way on how the that
no-cloning theorem is preserved during the process of
evaporation[5,24].\\
\hspace*{7.5mm} As it is suggested that a black hole might act as
a computer [15,16], we would like to calculate the speed of its
computation. For a black hole with the mass of 1 kg, it is
thermodynamical entropy can be given by the Bekenstein-Hawking
area law: $S=\frac{k_{B}}{4}l_{p}^{-2} A$, where $k_{B}$ is the
Boltzmann's constant, $l_{p}$ is the Planck length, and A is the
area of the black hole. By converting the entropy to the number of
bits of memory space available to the computer by using $S/k_{B}
ln2$, the amount of information that can be stored by the 1-kg
 the black hole can be given as $3.83\times 10^{16}$ bits. Since
 each teleportation in our case transferring 2 bits of message,
 the total operators that the 1-kg black hole should perform during its
 life-time is $1.915 \times 10^{16}$. Its life-time is given by $\tau\approx 10^{67}
 \frac{M}{M_{\odot}}$ (year) (${M_{\odot}}$ is the mass of the Sun), which is to say $\tau_{1kg}\approx
 10^{-18}$(second). Therefore, a 1-kg black hole can perform $\sim
 10^{34}$ operators per second. For a black hole with the mass of
 $M_{\odot}$, it can perform $\sim
 10^{3}$ operators per second. Comparing with our personal computer, the computation speed of such a Sun-like black hole
 is rather slow.\\
\hspace*{7.5mm} In summary, we have shown that  the annihilation
of the infalling particles and the collapsed matter within a black
hole can act as a measurement and transferring the information to
the outgoing Hawking radiation. We further demonstrated that H-M
proposal can be reached without nonlinear evolution of the
outgoing state, if the role of annihilation inside the horizon is
fully considered. The computation speed of a black hole is also
briefly
discussed.\\
 \textbf{Acknowledgements}\\ The work has been supported by the National
Natural Science Foundation of China under Grant No. 10273017.

\end{document}